\documentclass[aps,prb,twocolumn,showpacs]{revtex4-1}
\usepackage{graphicx}
\begin{document}

\title{Lattice dynamics of quasi-two-dimensional CdSe nanoplatelets \\
and their Raman and infrared spectra}

\author{Alexander I. Lebedev}
\email[]{swan@scon155.phys.msu.ru}
\affiliation{Physics Department, Moscow State University, 119991 Moscow,
Russia}

\date{\today}

\begin{abstract}
Phonon spectra of CdSe nanoplatelets (2--6~ML) with the zinc-blende structure
were calculated from first principles within the density-functional theory.
It turned out that the Lamb modes in nanoplatelets are in fact optical
rather than acoustic vibrations. Phonon spectra of the nanoplatelets
show the appearance of a large number of low-frequency modes inherited from
TA phonons in bulk CdSe. Calculations of the Raman spectra indicate a need
to revise the interpretation of available experimental data. The largest
contribution to the Raman spectra is provided by the quasi-Lamb modes with the
$A_1$~symmetry. The $B_2$~modes whose frequencies depend on the environment of
nanoplatelets and whose properties are closest to the properties of LO phonons
explain the results obtained in the ``nanoparticle-on-mirror'' geometry. The
features in Raman spectra previously attributed to surface optical (SO) modes
should be interpreted as a manifestation of lower-order quasi-Lamb $A_1$~modes.
Calculations of the infrared spectra find, in addition to the TO phonon line,
the appearance of intense lines from surface modes originating from terminating
F(Cl) atoms on the surface of nanoplatelets and true SO-modes.
\end{abstract}


\maketitle

\section{Introduction}
\label{sec1}

Vibrational spectroscopy is one of the most important methods for studying
materials. The lattice dynamics determines such physical properties of solids
as elastic moduli, sound velocity, heat capacity, thermal expansion,
thermal conductivity, and infrared absorption. In addition, the phenomena
associated with the electron-phonon interaction determine the mobility of
carriers in semiconductors and the changes in their electronic structure and
optical spectra with temperature.

To determine the vibration frequencies, infrared (IR) and Raman spectroscopy
as well as inelastic neutron scattering are used. Since the vibration
frequencies are very sensitive to the local chemical environment and the
structure of materials, the measurement of these frequencies and their
comparison with the calculated frequencies and atomic displacement patterns
can give detailed information on atomic and electronic structure of materials.

The physical properties of semiconductor nanoplatelets are qualitatively
different from those of bulk materials. In addition to the well-studied
quantum size effects, this also applies to the phonon spectra of these
quasi-two-dimensional systems. As was shown in a classical paper of
Lamb,~\cite{Lamb114}  in the spectra of acoustic vibrations of thin plates,
along with two acoustic modes with in-plane polarization, a new type of
acoustic vibrations arises, in which the atoms are displaced in the
out-of-plane direction and for which the quadratic dispersion law is
characteristic ($\omega \sim q^2$, where $q$ is the in-plane component
of wave vector). The peculiarities of optical vibrations in thin plates were
discussed in Refs.~\onlinecite{PhysRev.140.A2076,RepProgrPhys.33.149}.

The objects of this study are quasi-two-dimensional CdSe nanoplatelets with
the zinc-blende (ZB) structure. Like other cadmium chalcogenides, cadmium
selenide is currently an object of considerable interest due to unique
optical properties which are observed in nanoobjects of various shapes
and are controlled by the size effect and by creating of heterostructures.
In particular, CdSe colloidal nanoplatelets with a thickness of few
monolayers~\cite{JAmChemSoc.130.16504,NatureMater.10.936}
are characterized by atomically flat surfaces. Very homogeneous thickness
of the nanoplatelets results in very narrow luminescence bands. The giant
oscillator strength of the exciton transitions in
them~\cite{PhysRevLett.59.2337} results in very short luminescence decay times
(hundreds of picoseconds)~\cite{ACSNano.6.6751,NanoLett.13.3321} and high
absorption coefficients, which are much higher than those for quantum
dots.~\cite{JPhysChemC.119.20156}  Record narrow absorption and emission bands
resulting from the thickness-dependent exciton transitions make them promising
materials for the development of new types of light-emitting
devices,~\cite{AdvFunctMater.24.295,NanoLett.15.4611,ChemPhysLett.619.185}
lasers,~\cite{NatureNanotechnol.9.891,NanoLett.14.2772,NatureCommun.6.8513}
photodetectors,~\cite{NanoLett.15.1736} and other optoelectronic devices.

The literature data on Raman and IR spectra of CdSe nanoplatelets are not
too numerous.~\cite{SPIEProc.8807.88070A,PhysRevB.88.041303,
PhysRevLett.113.087402,Nanoscale.8.13251,Nanoscale.8.17204}
The main strong peak in the Raman spectra is explained by the light
scattering by LO phonons whose frequency varies with a change in the
nanoplatelet thickness as a result of the spatial confinement of the phonons.
The weak features observed on the low-energy side of this peak are explained by
surface optical (SO) phonons.~\cite{SPIEProc.8807.88070A,PhysRevB.88.041303,
Nanoscale.8.17204}  In Ref.~\onlinecite{SPIEProc.8807.88070A,PhysRevB.88.041303},
the thickness dependence of the frequencies of LO and SO modes was studied.
Under non-resonant excitation, the position of Raman peaks did not depend on
the thickness. Under resonant excitation (the excitation energy was close to
the energy of the exciton transition), the energy of the peaks decreased
noticeably (by 2.5--7.5~cm$^{-1}$) and their width slightly increased with
decreasing nanoplatelets thickness. This difference in behavior was explained
by preferential light scattering by LO($x$)~phonons (in the case of
non-resonant excitation) or by confined LO($z$)~modes (in the case of resonant
excitation). In Ref.~\onlinecite{PhysRevLett.113.087402}, Raman spectra from
individual nanoplatelets were obtained using the surface-enhanced
Raman scattering (SERS) in the ``nanoparticle-on-mirror'' (NPOM) geometry.
These spectra consisted of two overlapping lines, one of which, attributed
to the LO($z$) phonon, appeared at an energy by 7--10~cm$^{-1}$ lower than
the energy of the other line whose position was close to the energy
of the line measured in standard geometry and ascribed to LO($x$)
phonon. In Ref.~\onlinecite{Nanoscale.8.13251}, the low-frequency lines
associated with intrinsic mechanical vibrations of nanoplatelets were detected
in Raman spectra. The frequencies of these vibrations, however, were 40--50\%
below the calculated ones. This discrepancy was explained by the effect of
heavy molecules of oleic acid located on the surfaces of nanoplatelets
on the vibration frequency. In Ref.~\onlinecite{Nanoscale.8.17204},
Raman spectra of CdSe nanoplatelets were measured at low temperature
(35~K) and at different excitation energies. This measurements confirmed
the appreciable long-wave shift and broadening of the main line in the spectra
under resonant excitation and revealed a complicated structure of the SO mode.
In addition, in Ref.~\onlinecite{Nanoscale.8.17204} the first infrared reflection
spectra of CdSe nanoplatelets were obtained, and the features associated with
LO and TO phonons were observed.

In this work, the first-principles calculations of the lattice dynamics of CdSe
nanoplatelets are supplemented by modeling their Raman and infrared absorption
spectra to better understand what information can be obtained from these optical
studies. Based on the comparison of the results of calculations with available
experimental data, a new, more reasonable interpretation of these data is
proposed.

\section{Calculation details}

All calculations in this paper were performed within the density functional theory
using the \texttt{ABINIT} software package. In the calculations, the local
density approximation (LDA), plane-wave basis, and norm-conserving
pseudopotentials were used. The cut-off energy was 30~Ha (816~eV), the
integration over the Brillouin zone for the starting FCC lattice was carried
out on a 8$\times$8$\times$8 Monkhorst--Pack mesh. The relaxation of the unit
cell parameters and atomic positions was carried out until the Hellmann--Feynman
forces acting on the atoms became less than $5 \cdot 10^{-6}$~Ha/Bohr
(0.25~meV/{\AA}).

    \begin{figure}
    \centering
    \includegraphics{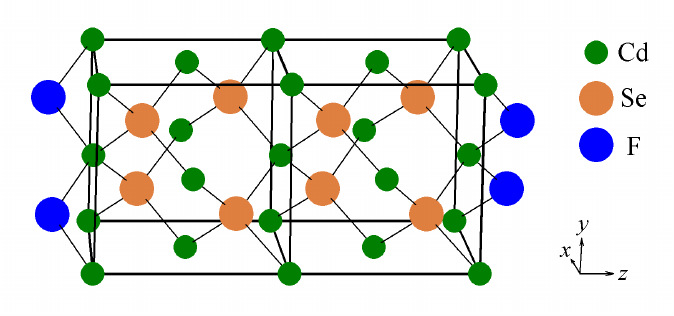}
    \caption{\label{fig1}The structure of CdSe nanoplatelet with a
    thickness of 4~ML.}
    \end{figure}

In accordance with modern concepts of the structure of CdSe nanoplatelets
with the ZB structure, the studied nanoplatelets were [001]-oriented slabs,
both surfaces of which ended in cadmium atoms (Fig.~\ref{fig1}). Since the
nanoplatelets contain one excess plane of Cd atoms, the extra charge of
$5s$~electrons introduced by these atoms requires electrical compensation.
In this paper, we considered the simplest method of charge compensation using
the surface F or Cl atoms. As shown in Ref.~\onlinecite{PhysRevB.95.165414},
energetically the most favorable position of these surface atoms are the bridge
positions in which the halogen atoms enter the positions of the missing selenium
atoms in the ZB structure (Fig.~\ref{fig1}).
Modeling of nanoplatelets was carried out using supercells to which a vacuum gap
of 20~{\AA} was added to isolate nanoplatelets from each other. The symmetry
of such a supercell is described by $P{\bar 4}m2$ space group. The thickness
of nanoplatelets varied from 2 to 6~monolayers (ML). When modeling the
nanoplatelets, 8$\times$8$\times$2 Monkhorst--Pack meshes were used for the
integration over the Brillouin zone. All the results presented below were
obtained on fully relaxed structures with stress-free surfaces.

Calculations of the phonon frequencies at the center and high-symmetry points
at the boundary of the Brillouin zone, the Born effective charges
$Z^*_{\alpha\beta}$, the high-frequency dielectric constant
$\epsilon^\infty_{\alpha\beta}$, the oscillator strengths
$S_{k,\alpha\beta}$, and the Raman susceptibility tensor were performed
using the formulas obtained from the density-functional perturbation
theory.~\cite{PhysRevB.55.10355,PhysRevB.71.125107}
Raman spectra of randomly oriented nanoplatelets were obtained using the
formulas given in Ref.~\onlinecite{PhysRevB.71.214307} by calculating the
rotational invariants of the Raman tensors and the Bose factors for 300~K.
Infrared spectra (imaginary part of the complex dielectric constant) were
calculated using the formula
$\epsilon_{\alpha\beta}(\omega) = \epsilon^\infty_{\alpha\beta} + (4\pi/\Omega) \sum_k S_{k,\alpha\beta}/[\omega_k^2 - \omega^2 - i\gamma\omega]$,
where the sum runs over all modes $k$, $\omega_k$ is the frequency of the
$k$th mode at the $\Gamma$ point, and $\Omega$~is the unit cell volume.
The calculated spectra were then averaged in order to obtain spectra of randomly
oriented nanoplatelets. Dispersion curves in the phonon spectrum were calculated
using the \texttt{anaddb} program (a part of the \texttt{ABINIT} package)
by interpolating the results of exact calculations of phonon frequencies on
a 8$\times$8 mesh of wave vectors (15~irreducible points in the Brillouin zone).

\section{Phonon spectra of the nanoplatelets}

    \begin{figure}
    \centering
    \includegraphics{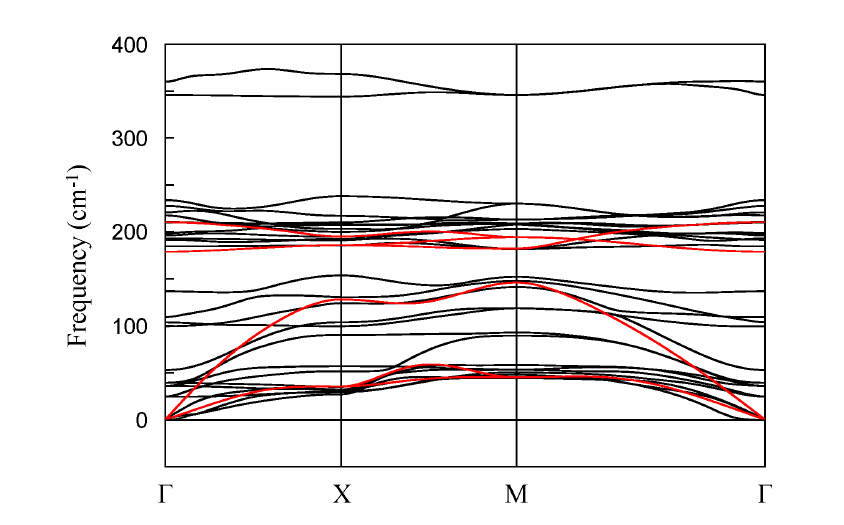}
    \caption{\label{fig2}Phonon spectra for CdSe nanoplatelet with a thickness
    of 3~ML terminated with F atoms (black lines) and bulk CdSe (red lines).}
    \end{figure}

The phonon spectrum of a typical quasi-two-dimensional CdSe nanoplatelet with
the ZB structure terminated by F~atoms is shown in Fig.~\ref{fig2}. In the
figure we present only the dispersion curves for the square (two-dimensional)
Brillouin zone because, as was confirmed by calculations on supercells,
the dispersion of the phonon modes in the out-of-plane direction is absent.

    \begin{figure}
    \centering
    \includegraphics{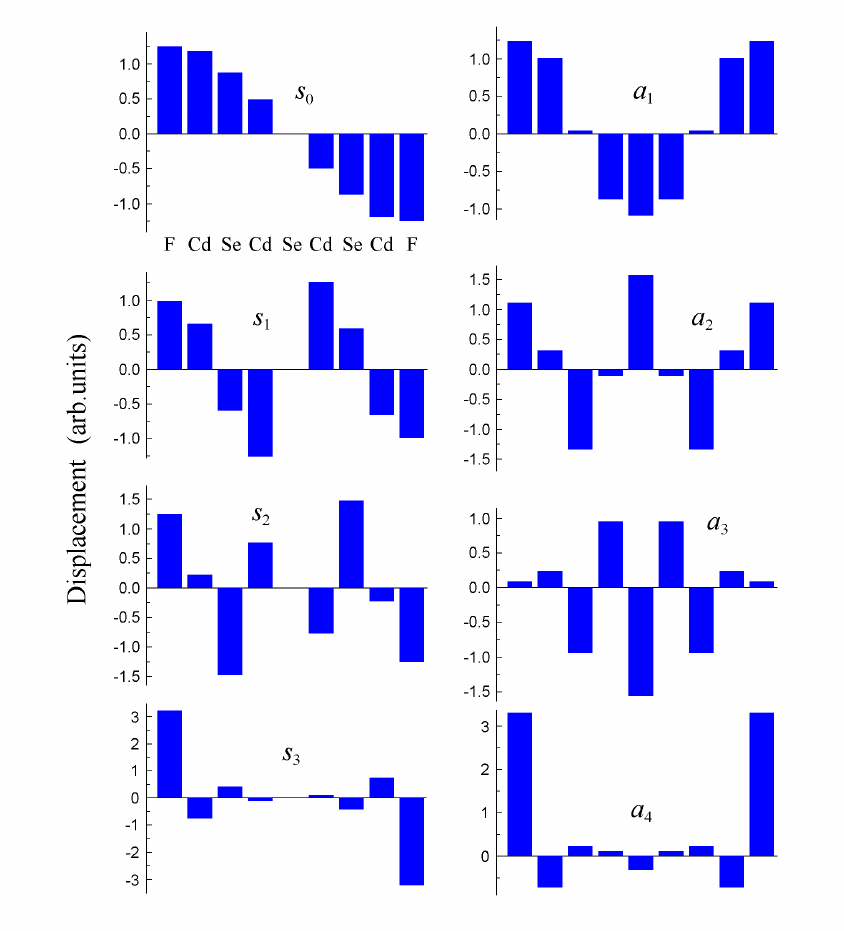}
    \caption{\label{fig3}The atomic displacement patterns of quasi-Lamb modes
    at the $\Gamma$~point for CdSe nanoplatelet with a thickness of 3~ML
    terminated with F atoms. The modes and their frequencies:
    $s_0$~--- 53~cm$^{-1}$, $a_1$~--- 100~cm$^{-1}$, $s_1$~--- 137~cm$^{-1}$,
    $a_2$~--- 196~cm$^{-1}$, $s_2$~--- 211~cm$^{-1}$, $a_3$~--- 221~cm$^{-1}$,
    $s_3$~--- 234~cm$^{-1}$, $a_4$~--- 228~cm$^{-1}$. The direction of the
    displacements is out-of-plane.}
    \end{figure}

We now discuss the character of phonon modes at the $\Gamma$~point for a typical
nanoplatelet with a thickness of 3~ML (nine atoms in the unit cell). After the
removal of three acoustic modes corresponding to the translation of the
nanoplatelet in three directions, 24~optical modes remain at the $\Gamma$~point.
In 16~of them, the atomic displacements occur in the plane of the nanoplatelet
($xy$), and in the other eight modes the displacements are out-of-plane ($z$~direction).
By the character of the displacement patterns, the latter eight modes are similar
to the Lamb modes (Fig.~\ref{fig3}). The symmetric Lamb modes~$s_i$%
    \footnote{To denote symmetric and antisymmetric Lamb modes, we use the
    notations $a_i$ and $s_i$, where $i$ indicates the order of the mode. The
    displacements of the atoms in these modes resemble the cosine function,
    with the $(i+1/2)$~wave periods for symmetric modes, and $i$~full periods
    for antisymmetric modes.}
(symmetry~$A_1$) have zero Born effective charge ($Z^* = 0$), whereas the
antisymmetric Lamb modes $a_i$ (symmetry~$B_2$) are characterized by a small
effective charge ($Z^*_{zz} = {}$0.070--1.039), which indicates the appearance
of polarization that accompanies vibrations. The appearance of the polarization
is not related to the piezoelectric properties of CdSe since it cannot arise
from the strain of the sample along the [001] direction. The polarization arises
as a result of the displacement of atomic planes carrying a small ionic
charge (the chemical bond in CdSe has a mixed covalent-ionic character). Because
of the appearance of a long-range dipole-dipole interaction, the corresponding
vibrational modes should be interpreted as \emph{optical} rather than acoustic
vibrations.%
    \footnote{In Lamb's original work, the waves in a thin plate were
    considered as purely acoustic waves in continuous media with zero-stress
    boundary conditions at the surfaces of the plate.}
In what follows we shall refer to them as quasi-Lamb modes. This feature of the
Lamb waves can also be characteristic of nanoplatelets made of other dielectric
materials.

    \begin{figure}
    \centering
    \includegraphics{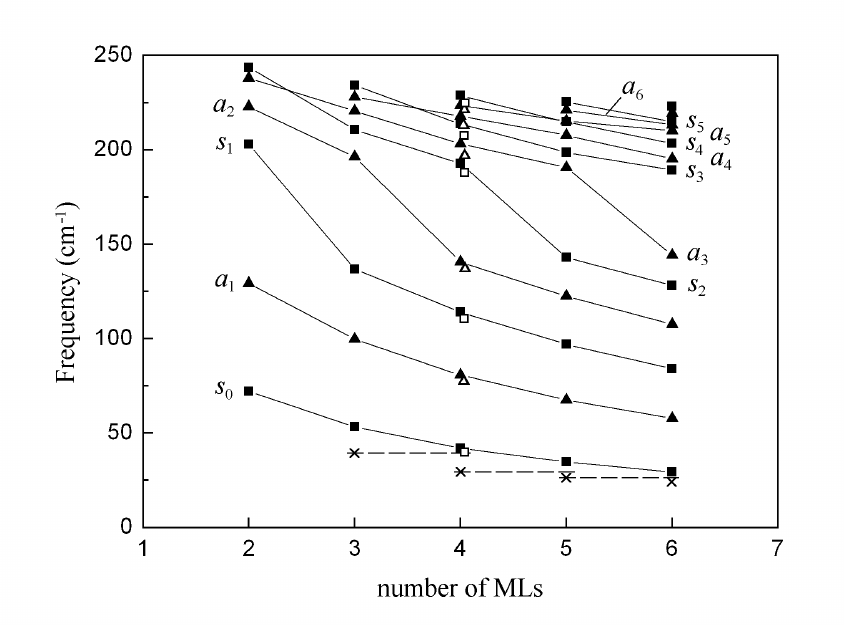}
    \caption{\label{fig4}Evolution of the frequencies of quasi-Lamb modes with
    a change in the thickness of CdSe nanoplatelets terminated by F atoms from
    2 to 6~ML. Open symbols show the frequencies for a nanoplatelet terminated
    by Cl atoms. Crosses are the experimental frequencies taken from
    Ref.~\onlinecite{Nanoscale.8.13251}.}
    \end{figure}

In order to clarify the relationship between quasi-Lamb modes and normal
modes in bulk CdSe, we performed the mode projection analysis by expanding the
eigenvectors of the quasi-Lamb modes in an orthogonal basis of normal modes of
bulk CdSe. This technique was first used for studying the vibrational spectra
of CdSe quantum dots in Ref.~\onlinecite{PhysRevB.59.2881}. In contrast to
quantum dots, the vibrational spectra of CdSe nanoplatelets are described by
good quantum numbers ($q_x$, $q_y$, indexes of the $s_i$ and $a_i$ Lamb modes),
and so the $z$~component of wave vector $q_z$ is the only parameter which can
be considered as an expansion parameter. The obtained results are presented in
the Supplemental Material.~\cite{Suppl} Due to the symmetry of nanoplatelets, the LA and
LO($z$) modes are the only modes which contribute to the $A_1$ and
$B_2$~quasi-Lamb modes. The largest contribution to the low-order $A_1$~modes
comes from LA phonons with $q_z \ne 0$, whereas the largest contribution
to the high-order $A_1$ modes is from LO($z$) phonons with $q_z \ne 0$. The
largest contribution to the $B_2$~mode $a_i$ whose index~$i$ is equal to the
number of MLs in the nanoplatelet is provided by $q = 0$ LO phonon of bulk
CdSe (this mode has the highest $Z^*_{zz}$ value). The largest contribution
to other $B_2$~modes are from the LA (low-order modes) or LO($z$) (high-order
modes) phonons with $q_z \ne 0$.

The frequencies of quasi-Lamb modes at the $\Gamma$~point as a function of the
thickness of nanoplatelets are shown in Fig.~\ref{fig4}. The figure does not
include results for the flexural $a_0$ mode whose frequency is always zero at
$q = 0$. It is seen that the frequencies of high-order quasi-Lamb modes are
grouped near the frequency of LO phonon in bulk CdSe (210~cm$^{-1}$ in our
calculations). When increasing the nanoplatelet thickness, the lower-order
modes ``split off'' from this group, and their frequencies rapidly decrease.
According to the mode projection analysis, the largest contribution to the
modes that are grouped around 210~cm$^{-1}$ is made by LO phonons. The fact
that the frequencies of these modes exceed the frequency of TO phonon in bulk
CdSe (179~cm$^{-1}$ in our calculations) indicates the existence of the
dipole-dipole interaction in both $A_1$ and $B_2$~quasi-Lamb modes in the
nanoplatelets. The replacement of the surface F atom by Cl results in a small
(2--6~cm$^{-1}$) lowering of the frequencies of all quasi-Lamb modes
(Fig.~\ref{fig4}).

The results shown in Fig.~{\ref{fig4}} can be used to determine the thickness
of nanoplatelets. As the frequencies of quasi-Lamb modes depend on the mass of
the surface atoms, and in the experimental nanoplatelets the surface atom
is usually oxygen whose mass is close to that of fluorine, then Fig.~{\ref{fig4}}
can be directly used to solve this problem. With this aim, we superimposed the
experimental data from Ref.~\onlinecite{Nanoscale.8.13251}  on our calculated
curves. As in Ref.~\onlinecite{Nanoscale.8.13251}, the observed vibration
frequencies are somewhat lower than the calculated ones, but if one assumes that
the nanoplatelet is actually by one monolayer thicker,%
    \footnote{To identify nanoplatelets, it is convenient to use the wavelength
    $\lambda$ of the first exciton peak in their absorption spectra. The authors
    of Ref.~\onlinecite{Nanoscale.8.13251} assumed the thickness to be 3~ML for
    nanoplatelets with $\lambda \approx 460$~nm, 4~ML for those with
    $\lambda \approx 515$~nm, and 5~ML for those with $\lambda \approx 550$~nm.}
the agreement becomes good. As for the model used in Ref.~\onlinecite{Nanoscale.8.13251},
one should note that a model with an additional mass located on the surface of
a nanoplatelet is not entirely adequate to the experiment since the thickness
of the oleic acid layer on the surface of the nanoplatelet is comparable with
the nanoplatelet thickness. Taking into account the difference between the sound
velocities in CdSe and oleic acid, the model~\cite{JAcoustSocAm.101.2649}
of a plate covered with a medium whose acoustic wave impedance is substantially
(about an order of magnitude) smaller than that of CdSe seems to be more adequate.

As was mentioned in Sec.~\ref{sec1}, one of the acoustic modes ($a_0$) in the
vibrational spectra of thin plates is purely flexural, and for small~$q$ its
dispersion law is $\omega \sim q^2$. Such a dispersion law can indeed be seen
in the phonon spectra of nanoplatelets in Fig.~\ref{fig2} in the vicinity of
the $\Gamma$~point.

    \begin{figure}
    \centering
    \includegraphics{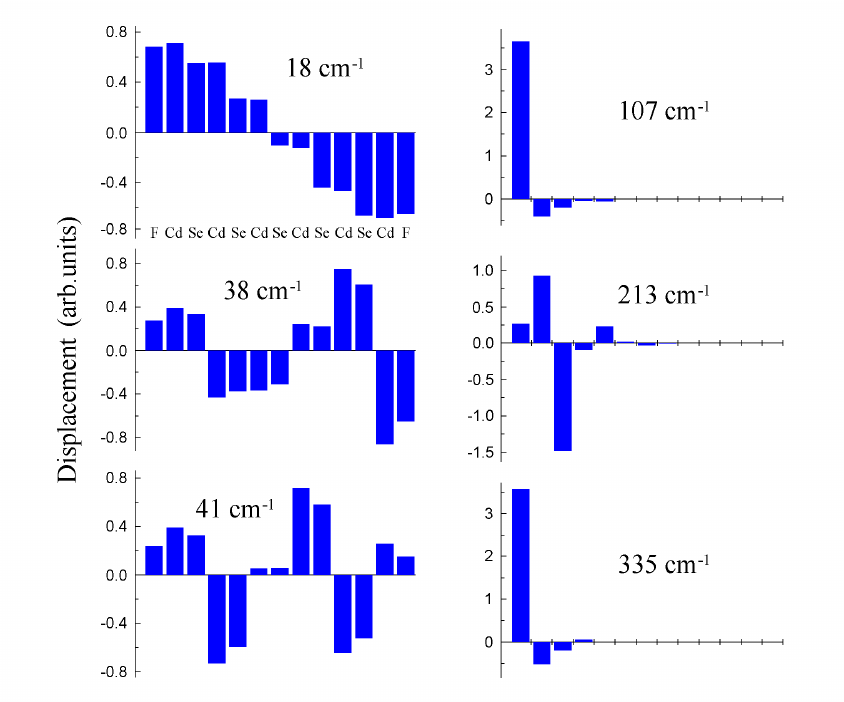}
    \caption{\label{fig5}Eigenvectors of several vibrational modes of the
    $E$~symmetry for CdSe nanoplatelet with a thickness of 5~ML terminated
    with F~atoms. The left panel shows modes with frequencies of 18~cm$^{-1}$,
    38~cm$^{-1}$, and 41~cm$^{-1}$. The right panel shows surface vibrational
    modes with frequencies of 107~cm$^{-1}$, 213~cm$^{-1}$, and 335~cm$^{-1}$.
    The atomic displacements are in the plane of the nanoplatelet.}
    \end{figure}

All optical modes with the $E$~symmetry and in-plane atomic displacements are
characterized by non-zero effective charge. When analyzing these modes, we see
a large number of modes that are absent in bulk CdSe and appear in the phonon
spectra of nanoplatelets in the frequency range 25--90~cm$^{-1}$ (Fig.~\ref{fig2}).
In order to understand their origin, we performed the mode projection
analysis of these modes (see the Supplemental Material~\cite{Suppl}). Due to the symmetry of
nanoplatelets, the TA and TO phonons are the only modes which contribute to the
$E$~modes. It turned out that in a nanoplatelet with a thickness of $n$~MLs,
there are $n$~modes with a dominant contribution from TA phonons of bulk CdSe
with $q_z \ne 0$, and the remaining modes have a dominant contribution from
TO phonons.

The characteristic ``doubling'' of atomic displacements, which is observed for
all acoustic-like $E$~modes on the left panel in Fig.~\ref{fig5}, and
the fact that the eigenvectors of these modes are directed along the [110] and
[1${\bar 1}$0] axes, support their relationship with TA phonons in bulk CdSe.
Indeed, when the atoms in the (001) planes in the ZB structure are displaced
along the indicated directions, two types of stresses of interatomic bonds arise
in adjacent planes: one generating elongation and shortening of bonds and the
other changing the angles between them without changing their lengths. Since
the energy cost necessary to change the angles between the bonds is much less
than that required to change their lengths, we immediately get a set of two
strongly different local ``shear moduli''. The ``doubling'' of atomic
displacements is a consequence of this difference: atomic layers bound by
stronger bonds experience smaller relative displacement. These pairs of atomic
layers are bound with other pairs by much weaker bending bonds, which result in
the low-frequency acoustic-like modes.

Next $n-1$ $E$~modes are the modes with frequencies in the range
181--194~cm$^{-1}$. The mode projection analysis shows that only the lowest
of these modes has a dominant contribution from $q = 0$ TO phonon of bulk
CdSe, whereas the modes with higher energies result from TO phonons with
$q_z \ne 0$.

Finally, the last group of the $E$~modes (right panel in Fig.~\ref{fig5})
can be interpreted as surface modes: the amplitude of atomic vibrations decays
very rapidly when moving deep into the nanoplatelet. The maximum displacement
is experienced by surface F~atoms or atoms of the first monolayer. The mode
projection analysis shows that they have a roughly equal contributions from all
TO phonons in the whole Brillouin zone. Modes with frequencies of 107 and
335~cm$^{-1}$, which have very similar displacement patterns, differ in
frequency by more than three times. In the first of these modes, the
displacements of F occur perpendicularly to their chemical bonds with Cd (the
bending mode), whereas in the second of them the displacement of fluorine causes
an elongation of one and shortening of the other of these bonds (the stretching
mode). This explains the difference in their frequencies. The frequencies
of these two modes significantly decrease (by 30--50\%) when replacing F~atoms
by Cl~atoms.

As concerns the mode with a frequency of 213~cm$^{-1}$, we associate it
with a so-called surface optical (SO) mode. According to the
theory,~\cite{RepProgrPhys.33.149} for a thin slab, the frequency of this
mode should exactly coincide with the frequency of bulk TO phonon.%
    \footnote{Surface modes with frequencies intermediate between those of TO
    and LO phonons can be observed only on curved surfaces: in quantum dots and
    quantum wires.}
Our calculations do indeed find such a surface $E$~mode (Fig.~\ref{fig5}).
In the nanoplatelet with a thickness of $n$~MLs, this mode always appears
as the highest-energy optical $E$~mode whose eigenvector qualitatively differs
from that of other optical $E$~modes. There are two reasons for the observed
upward shift of the frequency of the SO mode as compared to that of TO phonon:
(1)~according to the mode projection analysis, TO phonons from the whole
Brillouin zone contribute almost equally to this mode; (2)~there is a large
contribution from the terminating F~atom into this mode (the replacement of F by
Cl decreases the SO mode frequency by 13--15~cm$^{-1}$). In contrast to
quantum dots,~\cite{PhysRevB.59.2881} in nanoplatelets the SO mode has not
a mixed LO+TO character, but is a purely transverse mode.

We note that a similar picture with strong displacements of surface F~atoms
typical of surface modes can be seen for two highest-frequency quasi-Lamb
modes ($s_3$, $a_4$) in Fig.~\ref{fig3}. In the mode projection
analysis, these modes display a behavior that is different from the behavior
of both F surface modes and other quasi-Lamb modes. On the other hand, the
arguments against their surface character are: (1)~the frequencies of these
modes exhibit very little change (2--4~cm$^{-1}$) upon replacing the
terminating F~atoms by Cl~atoms and (2)~a sufficiently high displacement
amplitude is retained in the bulk of a nanoplatelet even for the thickest of
them. Nevertheless, we tend to classify these modes as surface modes, whose
$z$~displacements are strongly coupled with the LO-type displacements of atoms
in the bulk of the nanoplatelet due to the long-range dipole-dipole interaction.
This assumption let us propose the following simple picture that describes
the phonon spectrum of all nanoplatelets.

The unit cell of a CdSe nanoplatelet with a thickness of $n$~MLs contains
$2n+3$~atoms and produces $3(2n+3)$ vibrational modes. Three of them, the
flexural $a_0$ mode, one LA, and one TA phonons polarized in the $xy$~plane
are acoustic ones, the other $6(n+1)$~modes are optical. The first $6n$ modes
are produced by $n$ CdSe monolayers. These are: $n$~symmetric quasi-Lamb
$s_i$~modes ($i = 0\ldots n-1$), $n$~antisymmetric $a_i$~modes ($i = 1\ldots n$),
and $2n$~doubly degenerate optical $E$~modes, half of which are acoustic-like,
and the other half are optic-like (the TO-like mode with the highest energy
produces the so-called SO~mode). The remaining six modes, two doubly-degenerate
surface $E$~modes, $s_n$, and $a_{n+1}$ modes, are associated with
vibrations of two terminating F(Cl)~atoms.

\section{Interpretation of Raman and IR spectra}
\label{sec4}

    \begin{figure}
    \centering
    \includegraphics{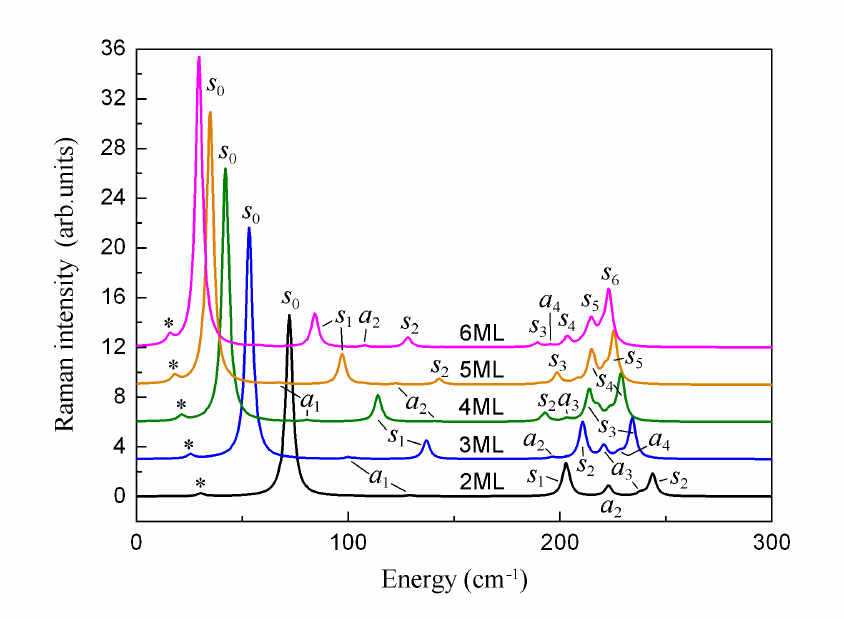}
    \caption{\label{fig6}Calculated Raman spectra for CdSe nanoplatelets with a
    thickness from 2 to 6~ML terminated with F~atoms. The asterisks show the
    most intense of the $E$~modes.}
    \end{figure}

The Raman spectra of CdSe nanoplatelets in the back-scattering geometry
calculated for non-resonant excitation conditions are shown in Fig.~\ref{fig6}.
In modeling the spectra, we intentionally used a small parameter of the Lorentz
broadening for a fine structure in the spectra could be seen; in the experiment,
these details can be less visible because of stronger damping of the vibrations.
It is seen that the largest contribution to the spectra is provided by the fully
symmetric quasi-Lamb $A_1$~modes $s_i$ and---to a much lesser degree---by
the high-order $B_2$~modes $a_i$, which are analogs of the LO~modes. The
relative contribution of the $B_2$~modes to the Raman scattering weakens as
the nanoplatelet thickness increases. From the modes with the $E$~symmetry,
a weak contribution to the Raman spectra is provided by the mode with
the lowest frequency (asterisks in Fig.~\ref{fig6}); the intensity of other
$E$~modes (including LO($x$)~mode) is by 1--2~orders of magnitude smaller.
In spite of a large spread in positions and intensities of individual lines near
210~cm$^{-1}$, the center of gravity of these lines is a rather smooth function
of the nanoplatelet thickness. It locates slightly above the LO phonon frequency
in bulk CdSe and slowly decreases with increasing nanoplatelet thickness.%
    \footnote{As follows from the calculation of Raman spectra in biaxially
    compressed bulk CdSe, this energy shift can be explained by the in-plane
    compression of the modeled nanoplatelets, which decreases with
    increasing nanoplatelet thickness.~\cite{PhysRevB.95.165414}}

The difference in intensities of the $A_1$ and $B_2$~Raman lines results
from the different symmetry of their corresponding Raman tensors,
    $$\left( \begin{array}{ccc}
    \alpha & 0 & 0 \\
    0 & \alpha & 0 \\
    0 & 0 & \gamma
    \end{array} \right)
    {\rm and}
    \left( \begin{array}{ccc}
    0 & \beta & 0 \\
    \beta & 0 & 0 \\
    0 & 0 & 0
    \end{array} \right).$$
Since the $\alpha$~values are usually two times larger than the $\beta$~values
and $\gamma \approx 0.1\alpha$, the isotropic invariant produced by the first of
these tensors results in much higher intensity of Raman lines than the symmetric
anisotropic invariant produced by the second tensor (for details see the
Supplemental Material~\cite{Suppl}).

The difference between the $A_1$ and $B_2$~modes is of fundamental character.
Although both of these modes can be expanded in a series of normal modes of
bulk CdSe, their properties are very different: only one of polar $B_2$~modes
transforms to the bulk polar LO phonon when increasing nanoplatelet thickness.
The $A_1$~modes, which can be considered as standing waves formed from two LO
phonons with $q_z \ne 0$, can appear only in the restricted geometry. The main
feature of the $A_1$~mode---it is non-polar and does not produce a long-range
electric field---has no corresponding solution in bulk systems. So, the
$A_1$~mode is a new type of vibration specific to quasi-two-dimensional
structures. Breathing modes
in quasi-0D systems and radial breathing modes in quasi-1D systems are other
examples of such modes. Thus, our calculations do not confirm the commonly
accepted interpretation of the main peak in Raman spectra of CdSe nanoplatelets
as a manifestation of LO phonon.

The results of our calculations show that the explanation of the observed
differences between non-resonant and resonant Raman
spectra~\cite{SPIEProc.8807.88070A,PhysRevB.88.041303}  may be incorrect, too.
In order for two lines of approximately equal amplitude to appear in the
resonant Raman spectrum, the intensity of the $E$~mode should be increased by
three orders of magnitude with respect to the intensity of the $A_1$~mode
with close frequency, which seems unlikely. A more natural explanation is
the change in the intensity of different $A_1$~modes, which form
a fine structure in the spectra, under resonant conditions. An experimental
confirmation of the existence of the fine structure in Raman spectra
is its observation in CdSe/CdS nanoheterostructures upon changing the
excitation energy.~\cite{Nanoscale.8.17204}

We now discuss the interpretation of weak features that are observed on the
low-energy side of the main Raman peak and are explained by surface optical
(SO) phonon. As was mentioned above, our calculations do indeed find
such a surface $E$~mode (mode with a frequency of 213~cm$^{-1}$ in
Fig.~\ref{fig5}). This mode should obviously be observed in the IR absorption
spectra (see below), but according to our calculations, in the Raman spectra
its intensity is three orders of magnitude smaller than the intensity of the
nearest strong line. This means that the features which have been explained
by SO phonon have a different origin. We believe that these features actually
result from other lines in the fine structure of the calculated Raman
spectra, first of all from the fully symmetric quasi-Lamb $A_1$~modes of
lower order. The fine structure of SO mode revealed in
Ref.~\onlinecite{Nanoscale.8.17204} supports this interpretation.
Obviously, the peaks attributed to the combination LO+SO in
Ref.~\onlinecite{Nanoscale.8.17204} should be interpreted as a second-order
scattering on the combination of the $A_1$~modes.

Long-wavelength shifts of Raman lines observed in the NPOM
geometry,~\cite{PhysRevLett.113.087402}  in our opinion, indicate the modes
that, in finite-size systems, create an electric field outside the sample and
therefore depend on the geometry of measurements and the dielectric constant
of surrounding medium. Among the modes existing in CdSe nanoplatelets
only the $B_2$~modes have such a property.

In order to verify the correctness of our reasoning, we calculated the mode
frequencies in several nanoplatelets with the same positions of atoms, but
with an increased thickness of the vacuum gap. It is obvious that the frequency
of the $B_2$~modes, for which the electric field is non-zero outside the
sample, should depend on the size of this gap because the electric field energy
in the gap is proportional to its thickness. It turned out that upon an increase
in the thickness of the vacuum gap by 6~{\AA} for the 3~ML nanoplatelet, the
frequencies of only two modes changed more than 0.01~cm$^{-1}$: these were
the $B_2$~modes---the $a_3$~mode (the frequency increment was 0.32~cm$^{-1}$)
and the $a_4$~mode (the frequency increment was 1.29~cm$^{-1}$). These are the
modes whose frequencies can change significantly when the environment of a
nanoplatelet changes. For the 5~ML nanoplatelet, a noticeable increase in the
frequency was also observed for two $B_2$ modes---the $a_5$~mode (the frequency
increment was 1.17~cm$^{-1}$) and the $a_6$~mode (the frequency increment was
0.51~cm$^{-1}$) for the same increase in the thickness of the vacuum gap.
Therefore, it is not surprising that when the gap in the NPOM geometry was
reduced, a decrease in the frequency of the $B_2$~mode was
observed.~\cite{PhysRevLett.113.087402}  The ``plasmonic amplification'' of the
intensity of this mode in the NPOM geometry is the reason for the appearance of
the second line in the Raman spectrum in addition to the $A_1$~line. Our
interpretation of the observed changes in the spectra differs from that proposed
in Ref.~\onlinecite{PhysRevLett.113.087402}. In addition, the authors are
not correct when considering the $B_2$~modes inactive in the standard Raman
geometry.

    \begin{figure}
    \centering
    \includegraphics{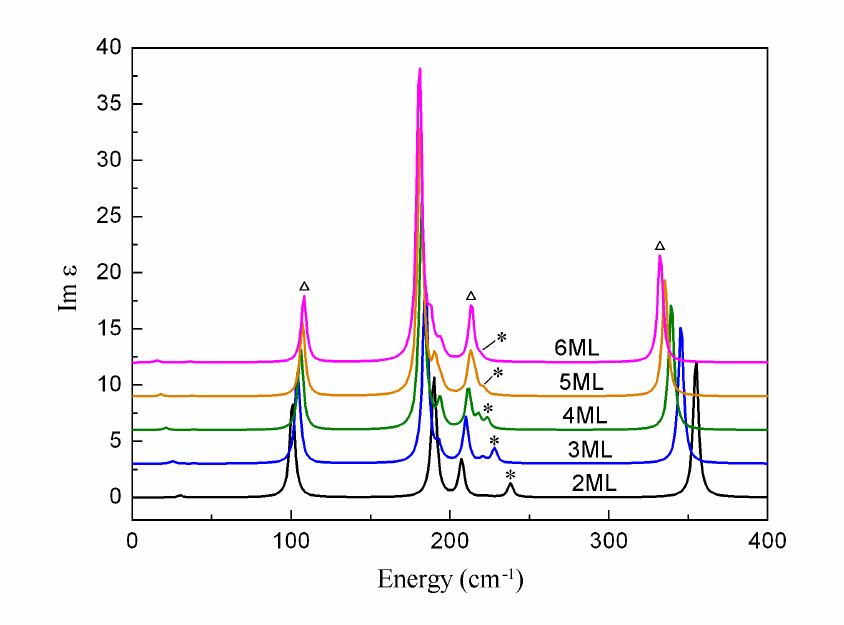}
    \caption{\label{fig7}Calculated infrared absorption spectra (Im~$\epsilon$)
    for CdSe nanoplatelets with a thickness from 2 to 6~ML terminated with
    F~atoms. Asterisks mark the $B_2$~modes, triangles mark the surface modes.}
    \end{figure}

Calculations of the infrared absorption spectra (Fig.~\ref{fig7}) show that
the largest contribution to them is provided by the $E$~mode with the highest
effective charge and the displacement pattern closest to that of TO phonon
in bulk CdSe. As the thickness of the nanoplatelet increases, the frequency of
this mode tends to the frequency of TO phonon in bulk CdSe. On the
high-energy side of this peak one can see a structure resulting from the
$E$~modes corresponding to TO phonons with $q_z \ne 0$. Three peaks of
medium intensity (triangles
in Fig.~\ref{fig7}) are related to the surface modes whose eigenvectors are
shown on the right panel in Fig.~\ref{fig5}. The true SO mode, which was
discussed above, can be clearly seen as a separate feature in the IR reflection
spectra.~\cite{Nanoscale.8.17204}  Finally, a very small contribution to the
spectra is provided by the quasi-Lamb $B_2$~modes (asterisks in Fig.~\ref{fig7}).
Although the dipole moments arising from the latter vibrations are directed
out-of-plane, we can observe them because nanoplatelets have a random
orientation with respect to the direction of light propagation. In summary,
we see that the IR spectra can provide important information on the
TO-like modes and the surface vibrations in nanoplatelets.

    \begin{figure}
    \centering
    \includegraphics{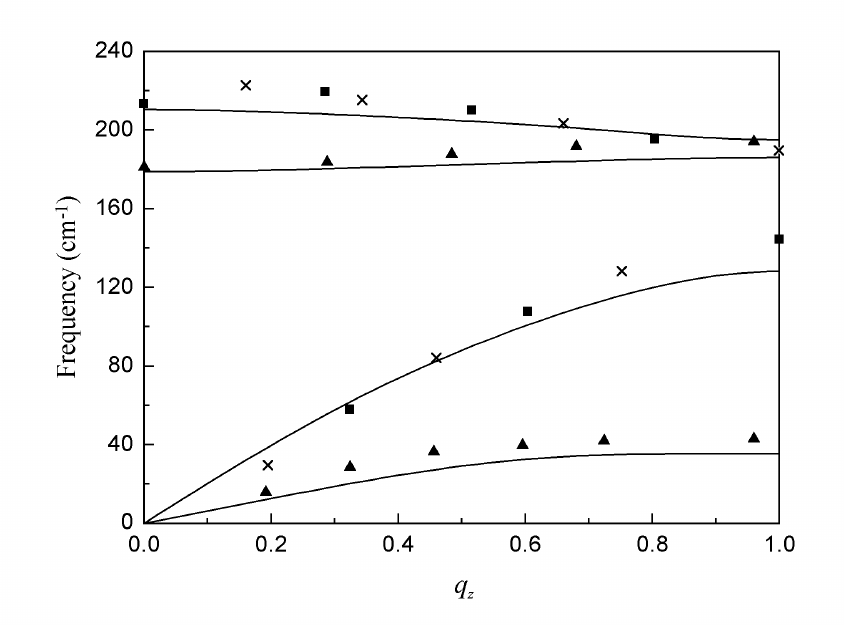}
    \caption{\label{fig8}Comparison between the frequencies of Raman-active
    modes for the 6~ML CdSe nanoplatelet and phonon spectrum of bulk CdSe. The
    $q_z$ values are the wave vectors that make the largest contribution to the
    modes according to the mode projection analysis. Triangles: data for the
    $E$~modes, squares: data for the $B_2$~modes, crosses: data for the
    $A_1$~modes.}
    \end{figure}

Lastly, a comment about the phonon confinement model~\cite{JRamanSpectrosc.38.604}
should be added. This model can indeed be regarded as a way for estimating
the shifts of LO and TO phonon frequencies in Raman spectra of nanostructured
materials. However, the phonon confinement model is usually concentrated on the
analysis of only these modes, whereas in this paper we have revealed much more
complex behavior resulting in the appearance of new vibrational modes
which are missing in the phonon confinement model, but are observable
in optical spectra. The frequencies of twenty-five Raman-active modes for
the 6~ML CdSe nanoplatelet are compared with the phonon spectrum of bulk CdSe
in Fig.~\ref{fig8}. It is seen that the frequencies of these modes are
qualitatively consistent with the frequencies of phonon modes of bulk CdSe taken
at $q_z$ vectors which provide the
largest contribution to the eigenvector of the corresponding mode. Nevertheless,
in some cases the frequencies can differ by 10--15~cm$^{-1}$, as for the
highest-energy mode which corresponds to the strongest line in the Raman spectrum.
We note that both the phonon dispersion curves and the mode frequencies presented
in the figure were calculated using the same technique. The origin of this
difference is that in our lattice dynamics calculations of nanoplatelets,
we used relaxed structures and calculated
the force constants from first principles individually for each atom,
thus avoiding the question of how to define ``the nanoplatelet thickness''.
The difference between our approach and the phonon confinement model
may be especially big when calculating Raman and IR spectra of
nanoheterostructures, in which the thickness the the outer layer can be as small
as 1~ML and where the phonon confinement model becomes inapplicable.

\section{Conclusions}

The first-principles calculations of the lattice dynamics of CdSe nanoplatelets
have shown that the Lamb modes in them are in fact optical, not acoustic
vibrations. In the spectra of optical vibrations of nanoplatelets, there appears
a large number of low-frequency modes inherited from TA phonons in bulk
CdSe. Calculations of the Raman spectra show that the largest contribution to
them is provided by the fully symmetric quasi-Lamb modes with the $A_1$~symmetry.
Antisymmetric quasi-Lamb modes with the $B_2$~symmetry, whose frequencies
depend on the environment of a nanoplatelet and whose properties are closest
to the properties of LO phonons, enable to explain the results of Raman studies
in the NPOM geometry. It is shown that weak lines in Raman spectra previously
attributed to surface optical (SO) phonons should be interpreted as a
manifestation of lower-order quasi-Lamb $A_1$~modes. Calculations of the
infrared spectra find, in addition to the TO phonon line, the appearance of
intense lines of surface modes originating from the atoms on the nanoplatelet
surface and from true SO modes. The results obtained in this work can be used
for a deeper understanding of the vibrational spectra of more complex
nanoobjects---nanoheterostructures and quantum dots. In quantum dots, for
example, the Raman lines may be associated with the breathing modes rather
than LO phonons, because the most intensive lines in the spectra are usually
the full-symmetry modes.

\medskip

\begin{acknowledgments}
The author is deeply grateful to R. B. Vasiliev for stimulating the author's
interest to nanoplatelets and fruitful discussions. This work was supported
by the Russian Foundation for Basic Research (Grant No. 16-29-11694).
\end{acknowledgments}

\providecommand{\BIBYu}{Yu}

\end{document}